\documentclass[prl,aps,10pt,twocolumn]{revtex4-1}

\usepackage{latexsym}
\usepackage{amstext}
\usepackage{amsmath}
\usepackage[normalem]{ulem}
\usepackage{subfigure}
\usepackage{epsfig, graphicx}
\usepackage{color}
\usepackage{amsfonts,bm,color}
\usepackage{verbatim}
\usepackage{float}
\usepackage{latexsym}
\usepackage[svgnames]{xcolor}
\usepackage{colortbl}
\usepackage{float}
\usepackage{wrapfig}
\usepackage{multirow}
\newcommand{\beq}{\begin{eqnarray}}
\newcommand{\eeq}{\end{eqnarray}}
\newcommand{\beqq}{\begin{eqnarray*}}
\newcommand{\eeqq}{\end{eqnarray*}}
\newcommand{\p}{\partial}

\newcommand{\eps}{\varepsilon}
\newcommand{\ETA}{\mbox{\boldmath$\eta$}}
\newcommand{\x}{\mbox{\boldmath$x$}}

\newcommand{\Q}{\mbox{\boldmath$Q$}}
\newcommand{\Pp}{\mbox{\boldmath$P$}}
\newcommand{\Bb}{\mbox{\boldmath$b$}}

\newcommand{\y}{\mbox{\boldmath$y$}}

\newcommand{\w}{\mbox{\boldmath$w$}}

\newcommand{\n}{\mbox{\boldmath$n$}}

\newcommand{\mb}[1]{ \mbox{\boldmath$#1$}}
\newcommand{\ds}{\displaystyle}

\begin{document}
\title{Extreme Narrow escape: shortest paths for the first particles to escape through a small window}
\author{K. Basnayake$^1$, A. Hubl$^1$, Z. Schuss$^2$,  D. Holcman$^1$}
\affiliation{$^1$ Ecole Normale Sup\'erieure, 75005 Paris, France}
\affiliation{$^2$ Tel-Aviv University, Tel-Aviv 69978, Israel.}
\date{\today}
\begin{abstract}
What is the path associated with the fastest Brownian particle that reaches a narrow window located on the boundary of a domain? Although the distribution of the fastest arrival times has been well studied in dimension 1, much less is known in higher dimensions. Based on the Wiener path-integral, we derive a variational principle for the path associated with the fastest arrival particle. Specifically, we show that in a large ensemble of independent Brownian trajectories, the first moment of the shortest arrival time is associated  with the minimization of the energy-action and the optimal trajectories are geodesics. Escape trajectories concentrate along these geodesics, as confirmed by stochastic simulations when an obstacle is positioned in front of the narrow window. To conclude paths in stochastic dynamics and their time scale can differ significantly from the mean properties, usually at the basis of the Smoluchowski's theory of chemical reactions.
\end{abstract}
\maketitle
Extreme statistics describe the properties of the shortest or longest arrival times in an ensemble of i.i.d. particles or their trajectories. The statistics of the shortest arrival time $\tau^{(n)}$ to a small target can be computed from the arrival times of a single particle \cite{extreme1,extreme3,extreme5,extreme6,extreme7}. When there are $n$ Brownian particles in a bounded domain $\Omega$, the shortest arrival time is defined by
\beq
\tau^{(n)}=\min (t_1,\ldots,t_n),
\eeq
where $t_i$ are the arrival times of the $n$ paths in the ensemble. The first moments of $\langle\tau^{(n)}\rangle$ were computed in dimension one \cite{Redner,majumdar2007brownian} and recently \cite{Kanishka2} in higher dimensions leading to the following asymptotic formulas:
\begin{align} \label{finalform}
({\bar{\tau}}^{(n)})^{dim 1} \approx& \frac{\delta^2}{4D\ln\left(\frac{n}{\sqrt{\pi}}\right)},\\ 
({\bar{\tau}}^{(n)})^{dim 2}\approx& \ds \frac{\ds \delta^2}{\ds 4 D \log\left(\frac{\pi
\sqrt{2}n}{8\log\left(\frac{1}{a}\right)}\right)},\\
({\bar{\tau}}^{(n)})^{dim 3}  \approx & \frac{\delta^2}{2D\sqrt{\log\left(\ds n\frac{4a^2}{\pi^{1/2}\delta^2}\right)}}, 
\end{align}
where $D$ is the diffusion coefficient, $\delta$ is the length of shortest ray from the initial point to the small exiting window and $n$ is the number of particles. \\
\begin{figure*}[http!]
	\center
	\includegraphics[scale=0.2]{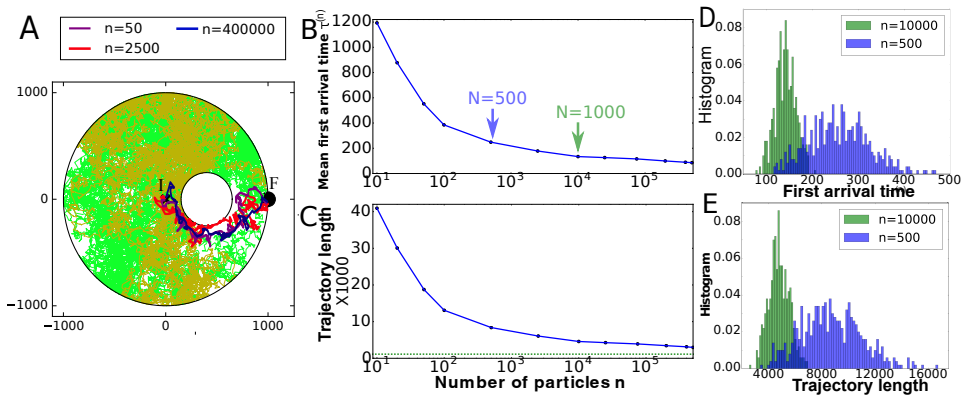}
	\caption{{\small {\bf Optimal paths associated to the fastest arrival time $\tau^{(n)}$ of $n$ i.i.d. Brownian trajectories} {\bf A.}. Brownian trajectories are initially positioned at point I, where they move inside a domain avoiding a round impenetrable obstacle. They can escape in a small window of size $\eps=0.01*R$ (R is the radius of the disk) located on the opposite side of point I. Classical escaping trajectories (green) are very different from the one associated to $\tau^{(n)}$ (bold face), concentrated as we shall see along geodesics (we just represented by symmetry the lower one). {\bf B.} MFPT associated to $\tau^{(n)}$   vs the number of particle $n$. {\bf C.} Trajectory length associated.{\bf D-E.} Distributions of times and lengths for $n=500$ and $10^4$.}}
\label{fig1}
\end{figure*}
The optimal search has many applications and the one for an ovule by spermatozoa moving inside an uterus was recently modeled using a coarse-grained rectilinear model for the sperms motion inside a cusp-like domain \cite{Yang}. Theory and simulations results showed that the trajectories of the fastest sperms to arrive to a narrow target are concentrated near the optimal trajectories of a control problem, which consists in minimizing the energy along the admissible paths. The situation was not much studied in the context of Brownian motion or stochastic processes in domains, especially when containing impenetrable obstacles. \\
We recall that the narrow escape problem (NEP) \cite{holcmanschuss2015} for the shortest arrival time is to find the probability density function (PDF) and the MFPT of $\tau^{(n)}$ for n=1. Here we focus on extreme trajectories associated with the fastest mean time $\langle\tau^{(n)}\rangle$  when $n$ is large. We show here that the paths used by the fastest particles to exit are concentrated near the optimal trajectories of a minimization problem, associated with the energy and are geodesics \cite{aubin2013some}. \\
This letter is organized as follows. We first discuss a path-integral formulation \cite{marchewka2000path,schuss2011nonlinear} for trajectories associated to the fastest arrival time. We then present a generic example of an extreme Narrow Escape Problem for the shortest arrival time among $n$ trajectories in two dimensions, where a disk obstacle is located exactly between the initial position and the exiting window. We estimate the splitting probability for the two possible ways out and show that the paths to the exit are concentrated near the shortest geodesics. Finally, we discuss how the shortest paths are strategically used in the context of stochastic processes and define the time scale of activation instead of the classical Smoluchowski's rate constant.\\
{\bf \noindent A variational principle associated to extreme statistics.}
The mean time $\langle\tau^{(n)}\rangle$ for n i.i.d. Brownian particles, when $n$ is large, can be expressed  as the complementary PDF of $t_1$,
\beq
\Pr\{\tau^{(n)}>t \} = {\Pr}^n\{t_{1}>t \},
\eeq
so that
{\small \beq\label{mfptmin}
\langle\tau^{(n)}\rangle=\int\limits\limits_0 ^{\infty}\Pr\{\tau^{1}>t\} dt = \int\limits_0 ^{\infty} \left[ \Pr\{t_{1}>t\} \right]^N dt.
\eeq
}
Here $\Pr\{t_{1}>t \}$ is the survival probability of a single particle prior to exiting at a target window (Fig. \ref{fig1}A). The survival probability is
\beq
\left[ \Pr\{t_{1}>t\} \right]^n=\exp \{ n\log \Pr \{ t_{1}>t \} \},
\eeq
where
\beq \label{surv}
\Pr\{t_{1}>t \} =\int\limits_{\Omega} p(\x,t)\,d\x.
\eeq
The transition probability density $p(\x,t\,|\,\y)$ is the solution of
\begin{align}\label{IBVP}
\frac{\p p(\x,t\,|\,\y)}{\p t} =&D \Delta p(\x,t\,|\,\y)\quad\hbox{\rm for } \x,\y \in \Omega,\\
p(\x,0\,|\,\y)=&\delta(\x-\y)\quad \hbox{\rm for } \x,\y \in\Omega\nonumber\\
\frac{\p p(\x,t\,|\,\y)}{\p \n} =&0\quad \hbox{\rm for } \x \in\p\Omega_r, \y\in\Omega\nonumber\\
p(\x,t\,|\,\y)=&0\quad \hbox{\rm for } \x \in \p\Omega_a,  \y\in\Omega,\nonumber
\end{align}
where the boundary $\p\Omega$ contains a small absorbing window
$\p\Omega_a$ and $\p\Omega_r=\p\Omega-\p\Omega_a$. Finally,
\begin{widetext}
\beq\label{mfptmin2}
\langle\tau^{(n)}\rangle=\int\limits\limits_0 ^{\infty}\exp \left\{ n \log  \int\limits_{\Omega} p(\x,t|\y)\,d\x\right\} dt =\int \tau_{\sigma} Pr\{ \hbox{ Path }\sigma \in S_n(\y), t<\tau_{\sigma}<t+dt \},
\eeq
\end{widetext}
where the ensemble $S_n(\y)$ is defined as all shortest paths starting at point $\y$ and exiting between time t and $t+dt$ from the domain $\Omega$ and selected among $n$. Our goal is to estimate $Pr\{ \hbox{ Path }\sigma \in S_n \}$ and in particular, to study whether or not the empirical stochastic trajectories in $S_n$ concentrate along the shortest paths starting from $\y$ and ending at the small absorbing window $\p \Omega_a$ under the condition that
$ \eps=\frac{|\p\Omega_a|}{|\p\Omega|} \ll 1$. To further describe the path ensemble $S_n(\y)$, we approximate path by discrete broken lines among a finite number of points. Using Bayes'rule,
{\small
\beqq
Pr\{ \hbox{ Path }\sigma \in S_n(\y)| t<\tau_{\sigma}<t+dt \}=\\\sum_{0}^{\infty}
Pr\{ \hbox{ Path }\sigma \in S_n(\y)|m, t<\tau_{\sigma}<t+dt \} Pr\{ m \mbox{ steps}\}
\eeqq
}
where $Pr\{ m \mbox{ steps}\}$ is the probability that a path exits in $m-$discrete time steps. When there are $n-$paths, a path made of broken lines (random walk with step $\Delta t$) can be expressed using Wiener path-integral. Indeed, for a stochastic process
\beq \label{stochastic}
d\x= \mb{a}({\x})dt +\mb{b}(\x) d\w
\eeq
the probability density at $\x,t$ is
\begin{widetext}
{\small
\beq
\Pr\Big{\{}{\x}_N(t_{1,M})\in\Omega,{\x}_N(t_{2,M})\in\Omega,\dots, {\x}_M(t)=\x, t\leq T\leq t+\Delta t\,|\,\x(0)=\y\Big{\}}=
\Bigg{[}\int\limits_{\Omega} \cdots
\int\limits\limits_{\Omega}\,\prod_{j=1}^{M} \frac{d{\y}_j}{\sqrt{(2\pi \Delta t)^n\det\mb{\sigma}(\x)(t_{j-1,M}))}}&& \nonumber\\
\exp \Bigg{\{} -\frac{1}{2\Delta t}
\left[\mb{\y}_j-\x(t_{j-1,N})- \mb{a}({\x}(t_{j-1,N}))\Delta t
\right]^T\mb{\sigma}^{-1}(\x(t_{j-1,N})) \left[{\y}_j-\x(t_{j-1,N})-\mb{a}(\x(t_{j-1,N}))\Delta t
\right]\Bigg{\}}&& ,\label{Wiener1}
\eeq}
\end{widetext}
where $\mb{\sigma}=\frac{\mb{b}^t\mb{b}}{2}$, $\Delta t=t/M,\  t_{j,N}=j\Delta t,\ \x(t_{0,N})=\y$ and $\mb{\y}_j=\x(t_{j,N})$ in the product and $T$ is the exit time in the narrow absorbing window $\p\Omega_a$.  We consider the case of a pure Brownian motion where $\mb{\sigma}=D$ is a constant and $\mb{a}=0$.
In the limit $n$ and $m$ large, we get
\begin{widetext}
{\small
\beqq
Pr\{ \hbox{ Path }\sigma \in S_n(\y)|m, t<\tau_{\sigma}<t+dt \}= \left(\int\limits_{\mb{\y}_0=\y} \cdots \int\limits_{\mb{\y}_j \in\Omega}\int\limits_{\mb{\y}_n\in\p\Omega_a} \frac{1}{(4D\Delta t)^{dm/2}}\prod_{j=1}^{m} \exp \Bigg{\{} -\frac{1}{4D\Delta t} \left[|\mb{\y}_j-\mb{\y}_{j-1})|^2 \right]\Bigg{\}}\right)^n \\
\approx \left(\frac{1}{(4D\Delta t)^{dm/2}}\right)^n  \int_{\x}\exp \Bigg{\{}-n\int\limits_0^{m \Delta t} \dot{\x}^2ds \Bigg{\} }{\mathcal D} (\x) \nonumber
\eeqq
}
\end{widetext}
where ${\mathcal D} (\x)$ is the limit Wiener measure \cite{schuss2011nonlinear}. Thus, we can re-write the first moment of $\tau^{(n)}$ by conditioning on the number of steps before exit
\begin{widetext}
\beq \label{condi}
\langle\tau^{(n)}\rangle& =&\sum_0^{\infty} (m\Delta t) Pr\{ \mbox{ shortest path among }  n | m \mbox{ steps } \} Pr\{ m \mbox{ steps}\} \\
&\approx& \int_0^{\infty} t  \int_{\x}\exp \Bigg{\{}-n\int\limits_0^{t} \dot{\x}^2ds \Bigg{\} } Pr\{ m \mbox{ steps}\} {\mathcal D} (\x). \nonumber
\eeq
\end{widetext}
For large $n$, the minimization of the exponent in eq. \ref{condi} is obtained by the Laplace's method, which allows selecting the paths for which the energy functional is minimal, that is
\beq\label{min}
E=\min_{X\in \mathcal P_t}\int\limits_0^T \dot{\x}^2ds,
\eeq
where the integration is taken over the ensemble of regular paths $\mathcal P_t$ inside $\Omega$ starting at $\y$ and exiting in $\p \Omega_a$, defined as
{\small \beqq
\mathcal P_T= \{ P(0)=\y, P(T)\in \p \Omega_a \hbox{ and } P(s) \in \Omega \hbox{ and }   0\leq s\leq T\}.
\eeqq
}
The Euler-Lagrange equations for the extremal problem \ref{min} are  the classical geodesics between $\y$ and a point in the narrow window $\p \Omega_a$. To conclude, in the limit of large $n$, the Brownian paths contributing to the first moment of $\langle\tau^{(n)}\rangle$ are concentrated on the geodesics, solution of the variational problem \ref{min}. Fig. \ref{fig1}A show the paths associated to increasing $n$ concentrated around the shortest geodesics. Fig. \ref{fig1}B-C show how the mean first arrival time for the fastest and the length of the associated path decay with n. Finally as n increases the associated histogram concentrates, as shown in Fig. \ref{fig1}B-C.\\
We shall now focus on an asymmetric example of Brownian escape when a round obstacle is positioned between the initial point and the narrow escaping window (fig. \ref{fig2}). In absence of such obstacle, the optimal path is the ray, used to construct the asymptotic solution \cite{Kanishka2}.\\
{\bf \noindent Optimal paths for an asymmetric escape between $\y$ and $\p \Omega_a$.}
\begin{figure*}[http!]
\center \includegraphics[scale=0.2]{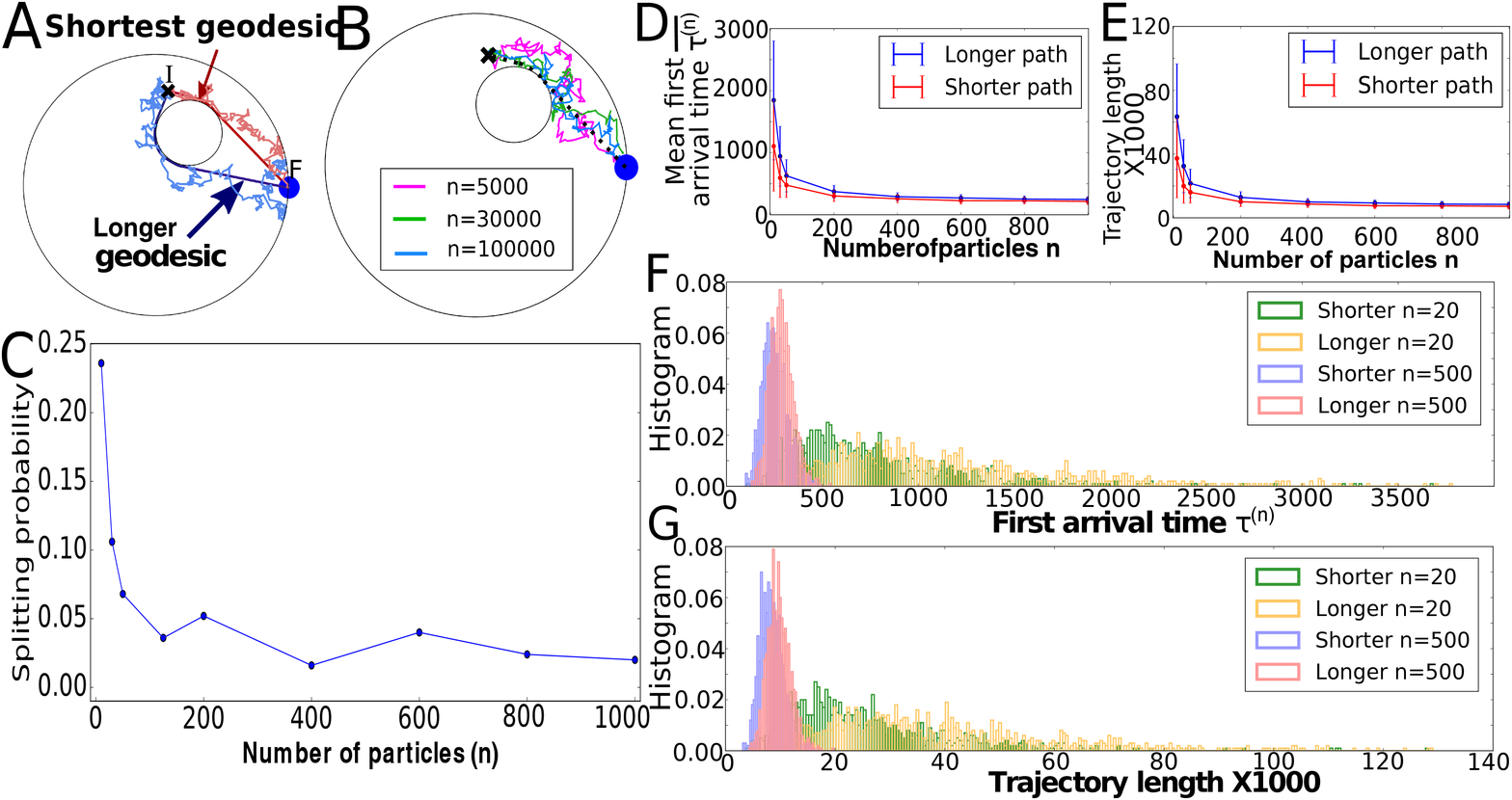}
\caption{{\small {\bf Optimal paths associated to the fastest arrival time of Brownian trajectories for an asymmetric initial point} {\bf A-B.} Shortest path compared to the geodesics and concentration of the path for increasing $n$.{\bf C.} Splitting probability computed empirically as the ratio of arrival times from above and below. {\bf D-F.} Mean arrival time and mean length of the trajectories for the fastest for the short and longer geodesics. {\bf G-H.} histogram of the arrival time and the length for $n=20$ and $500$.   }}\label{fig2}
\end{figure*}
We run stochastic simulations to determine the path associated with fastest arrival time among $n$ trajectories (Fig. \ref{fig1}). The domain $\Omega$ contains an obstacle positioned between the initial $\y$ and the narrow absorbing window $\p \Omega_a$.  As n increases, the trajectories associated with $\langle\tau^{(n)}\rangle$ concentrate near the geodesic ( $n=10^4, 10^5$ and $10^6$). In the symmetric case (Fig. \ref{fig1}), there two identical shortest paths which consists of straight line from $\y$ to the tangent of the disk, then an arc along the disk and finally a straight line from the tangent of the disk to the center of $\p \Omega_a$. When the initial point  $\y$  is not on the axis of symmetry, the two geodesics consists of one shorter than the other (Fig. \ref{fig2}).  In that case,  to estimate the splitting probability between trajectories associated with  $\tau^{(n)}$, we use stochastic simulations (Fig. \ref{fig2}C) to estimate first the mean arrival time from above $\tau_a^{(n)}$ and below $\tau_b^{(n)}$ (Fig. \ref{fig2}A-B) and we compute the ratio of one time to the total time $P_1=\frac{\langle\tau_a^{(n)}\rangle}{\langle\tau_a^{(n)}\rangle+\langle\tau_b^{(n)}\rangle}$. To add The splitting probability
The mean and the histograms associated with the arrival time and the length of the shortest Brownian path along the longer and shorter geodesic show that the statistics is not very different in the two cases, but the main effect is the concentration of the pdfs as n increases  (Fig. \ref{fig2}D-H).\\
\begin{figure}[http!]
	\center
	\includegraphics[scale=0.15]{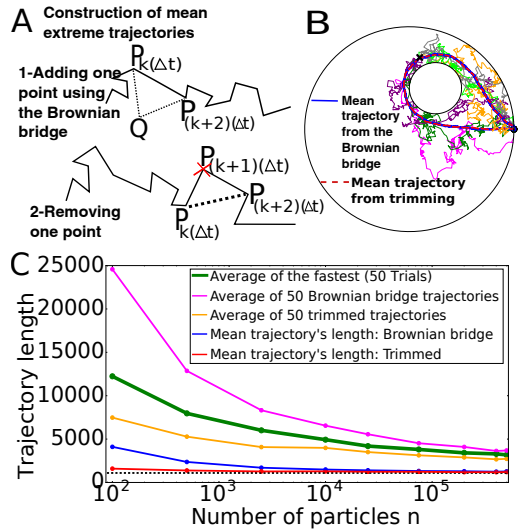}
    \caption{{\small {\bf Computing the mean escape paths.} {\bf A.} Schematic procedure to add or eliminate one point from a trajectory arriving with the fastest time among n particles.  {\bf B.} Computing the mean maximum and minimum length trajectories (thick line) from the ensemble of fastest trajectories. {\bf C.} Mean average, longest and shortest trajectories for increasing n.}}\label{fig3}
\end{figure}
{\bf \noindent Computing the mean shortest path associated to $\langle\tau_a^{(n)}\rangle$}
What is the mean path associated to $\langle\tau_a^{(n)}\rangle$? From empirical paths, we examine the path defined by piecewise segments connecting two neighboring points generated by a finite $\Delta t-$time step random walks. To each realization $k$, we chose the optimal path $\gamma^1_k$ that generate the ensemble $\Gamma_p=(\gamma^1_n,..\gamma^p_n)$  associated with the shortest arrival times $\tau^1_n,..\tau^p_n$, each are selected as the fastest time among $n$ random walks.  The mean time is computed from the empirical sum limit
\beq
\langle\tau_a^{(n)}\rangle \approx \lim_{p\rightarrow \infty }\frac{1}{p}\sum_{k=1}^p \tau^k_N.
\eeq
However, constructing an average path is not straightforward. Indeed the number of points for each individual trajectory in $\Gamma_p$ are different, but we can construct the mean shortest trajectory by selecting the path with the minimum number of points $n_{min}$ and then for any other longer trajectories, we remove points chosen uniformly along the path, so that all resulting trajectories have now the same number of minimum points. The mean shortest trajectory $\Gamma_{min}$ is computed by averaging the position at each step
\beq
\Gamma_m( k\Delta t) =\frac{1}{p} \sum_{1}^{p} \tilde\gamma^k_N (k\Delta t).
\eeq
and thus contains exactly $n_{min}$ points. It is associated to the mean of the fastest time $\langle\tau_a^{(n)}\rangle$.  The trajectory is obtained by joining any two neighboring points by a straight line (Fig. \ref{fig3}A-B). \\
It is also possible to compute the mean trajectory, not necessarily for the shortest path. To construct a longer path, we can construct a mean trajectory with $n_{min}+1$ points as follows: we apply the procedure mentioned above to all longer paths, however, we need to add one point for the shortest path. For that purpose, we use the discrete Brownian bridge at step $\Delta t$. We then chose a point  $P(k_s\Delta t)$ uniformly distributed in the sequence of points $(P(\Delta t),..P((n_{min}+1)\Delta t)$ for a path with $n_{min}+1$ points. Then we insert a point $Q_b$ between $P(k_s\Delta t)$  and $P((k_s+1)\Delta t)$ (Fig. \ref{fig3}A), so that the jump from $P(k_s\Delta t)$ to $Q_b$ is conditioned on the next event to be at $P((k_s+1)\Delta t)$ during the next time step $\Delta t$. The Brownian bridge is defined by
\beq
\Q_b=\frac{\Pp(k_s\Delta t)+\Pp((k_s+1)\Delta t)}{2} +\sqrt{2D \frac{\Delta t}{2}}\ETA, \nonumber
\eeq
where $\ETA$ is the a Gaussian variable of variance one and mean zero.  This procedure can be iterated for any number of points, thus generating a piecewise constant path with any given number of points. We use this procedure to obtain the mean trajectory with a number of points $n_p$ between the minimum $n_{min}$ and maximum $n_{max}$ number for trajectory in the ensemble $\Gamma_p$. Examples are shown in (Fig. \ref{fig3}B-C).\\
{\noindent \bf Conclusion.}
The exit paths associated with the shortest time among $n$ independent Brownian motion, in the limit of $n$ large, concentrate near the shortest geodesic. The present approach can be generalized to the exit of n i.i.d  stochastic processes \ref{stochastic} and in that case, the optimal trajectory are the ones obtained from the associated control problem $\min \int_0^T |\dot{\x}(s)-\Bb(\x(s))|^2ds$ \cite{freidlin1998random}. In addition, the present numerical simulations confirms the $log N$ decay of the extreme statistics NET formula, thus we can conjecture that even with obstacles, the mean first exit time is given by
\beq \label{finalform2}
\bar{\tau}^{d2}_N\approx&  \ds \frac{\ds \min (l_{g1}^2,..l_{gn}^2)}{\ds 4 D \log\left(\frac{\pi \sqrt{2}N}{8\log\left(\frac{1}{\eps}\right)}\right)},
\eeq
where the $l_g$ are the length of the shortest geodesic.\\
The present result shows that redundancy (having many copies of the same signaling molecule) defines the time of activation by selecting the shortest time through the shortest path from the source of production to the located of the target, located at a certain distance away. This principle is likely to be generic in cellular transduction \cite{Fain}, due to the organization of many organelles that constitute impenetrable obstacles. Finally, in the absence of a steady state concentration, the present approach revealed that the rate of activation through is not given by the forward rate of the Smoluchowski formula \cite{JPA2016-HS}, showing that the classical chemical reaction theory based on the mass-action law or the Gillespie'algorithm based on the forward rate are not appropriate to study transient activation with small targets.\\
Another consequence of the present approach is a warning about the method of computation of the effective diffusion coefficient by homogenization of large obstacles. This computation is based on the NET and thus on the NET \cite{holcmanschuss2015}.  However, it is not relevant to compute the MFPT of the fastest among n Brownian particles from an effective diffusion coefficient, rather the MFPT should be computed directly from the molecular scale by accounting for the shortest geodesic, as shown here. 
\normalem
\bibliographystyle{apsrev4-1} 

\bibliography{biblio4,biblio7}

\begin{thebibliography}{16}%
\makeatletter
\providecommand \@ifxundefined [1]{%
 \@ifx{#1\undefined}
}%
\providecommand \@ifnum [1]{%
 \ifnum #1\expandafter \@firstoftwo
 \else \expandafter \@secondoftwo
 \fi
}%
\providecommand \@ifx [1]{%
 \ifx #1\expandafter \@firstoftwo
 \else \expandafter \@secondoftwo
 \fi
}%
\providecommand \natexlab [1]{#1}%
\providecommand \enquote  [1]{``#1''}%
\providecommand \bibnamefont  [1]{#1}%
\providecommand \bibfnamefont [1]{#1}%
\providecommand \citenamefont [1]{#1}%
\providecommand \href@noop [0]{\@secondoftwo}%
\providecommand \href [0]{\begingroup \@sanitize@url \@href}%
\providecommand \@href[1]{\@@startlink{#1}\@@href}%
\providecommand \@@href[1]{\endgroup#1\@@endlink}%
\providecommand \@sanitize@url [0]{\catcode `\\12\catcode `\$12\catcode
  `\&12\catcode `\#12\catcode `\^12\catcode `\_12\catcode `\%12\relax}%
\providecommand \@@startlink[1]{}%
\providecommand \@@endlink[0]{}%
\providecommand \url  [0]{\begingroup\@sanitize@url \@url }%
\providecommand \@url [1]{\endgroup\@href {#1}{\urlprefix }}%
\providecommand \urlprefix  [0]{URL }%
\providecommand \Eprint [0]{\href }%
\providecommand \doibase [0]{http://dx.doi.org/}%
\providecommand \selectlanguage [0]{\@gobble}%
\providecommand \bibinfo  [0]{\@secondoftwo}%
\providecommand \bibfield  [0]{\@secondoftwo}%
\providecommand \translation [1]{[#1]}%
\providecommand \BibitemOpen [0]{}%
\providecommand \bibitemStop [0]{}%
\providecommand \bibitemNoStop [0]{.\EOS\space}%
\providecommand \EOS [0]{\spacefactor3000\relax}%
\providecommand \BibitemShut  [1]{\csname bibitem#1\endcsname}%
\let\auto@bib@innerbib\@empty
\bibitem [{\citenamefont {Sokolov}\ \emph {et~al.}(2005)\citenamefont
  {Sokolov}, \citenamefont {Metzler}, \citenamefont {Pant},\ and\ \citenamefont
  {Williams}}]{extreme1}%
  \BibitemOpen
  \bibfield  {author} {\bibinfo {author} {\bibfnamefont {I.~M.}\ \bibnamefont
  {Sokolov}}, \bibinfo {author} {\bibfnamefont {R.}~\bibnamefont {Metzler}},
  \bibinfo {author} {\bibfnamefont {K.}~\bibnamefont {Pant}}, \ and\ \bibinfo
  {author} {\bibfnamefont {M.~C.}\ \bibnamefont {Williams}},\ }\href@noop {}
  {\bibfield  {journal} {\bibinfo  {journal} {Physical Review E}\ }\textbf
  {\bibinfo {volume} {72}},\ \bibinfo {pages} {041102} (\bibinfo {year}
  {2005})}\BibitemShut {NoStop}%
\bibitem [{\citenamefont {Yuste}\ and\ \citenamefont
  {Lindenberg}(1996)}]{extreme3}%
  \BibitemOpen
  \bibfield  {author} {\bibinfo {author} {\bibfnamefont {S.~B.}\ \bibnamefont
  {Yuste}}\ and\ \bibinfo {author} {\bibfnamefont {K.}~\bibnamefont
  {Lindenberg}},\ }\href@noop {} {\bibfield  {journal} {\bibinfo  {journal}
  {Journal of statistical physics}\ }\textbf {\bibinfo {volume} {85}},\
  \bibinfo {pages} {501} (\bibinfo {year} {1996})}\BibitemShut {NoStop}%
\bibitem [{\citenamefont {Majumdar}\ and\ \citenamefont
  {Pal}(2014)}]{extreme5}%
  \BibitemOpen
  \bibfield  {author} {\bibinfo {author} {\bibfnamefont {S.~N.}\ \bibnamefont
  {Majumdar}}\ and\ \bibinfo {author} {\bibfnamefont {A.}~\bibnamefont {Pal}},\
  }\href@noop {} {\bibfield  {journal} {\bibinfo  {journal} {arXiv preprint
  arXiv:1406.6768}\ } (\bibinfo {year} {2014})}\BibitemShut {NoStop}%
\bibitem [{\citenamefont {Chou}\ and\ \citenamefont
  {D’orsogna}(2014)}]{extreme6}%
  \BibitemOpen
  \bibfield  {author} {\bibinfo {author} {\bibfnamefont {T.}~\bibnamefont
  {Chou}}\ and\ \bibinfo {author} {\bibfnamefont {M.}~\bibnamefont
  {D’orsogna}},\ }\href@noop {} {\bibfield  {journal} {\bibinfo  {journal}
  {First-Passage Phenomena and Their Applications}\ }\textbf {\bibinfo {volume}
  {35}},\ \bibinfo {pages} {306} (\bibinfo {year} {2014})}\BibitemShut
  {NoStop}%
\bibitem [{\citenamefont {Schehr}\ \emph {et~al.}(2014)\citenamefont {Schehr},
  \citenamefont {Majumdar}, \citenamefont {Oshanin},\ and\ \citenamefont
  {Redner}}]{extreme7}%
  \BibitemOpen
  \bibfield  {author} {\bibinfo {author} {\bibfnamefont {G.}~\bibnamefont
  {Schehr}}, \bibinfo {author} {\bibfnamefont {S.~N.}\ \bibnamefont
  {Majumdar}}, \bibinfo {author} {\bibfnamefont {G.}~\bibnamefont {Oshanin}}, \
  and\ \bibinfo {author} {\bibfnamefont {S.}~\bibnamefont {Redner}},\
  }\href@noop {} {\bibfield  {journal} {\bibinfo  {journal} {First-Passage
  Phenomena and Their Applications}\ }\textbf {\bibinfo {volume} {35}},\
  \bibinfo {pages} {226} (\bibinfo {year} {2014})}\BibitemShut {NoStop}%
\bibitem [{\citenamefont {Meerson}\ and\ \citenamefont
  {Redner}(2015)}]{Redner}%
  \BibitemOpen
  \bibfield  {author} {\bibinfo {author} {\bibfnamefont {B.}~\bibnamefont
  {Meerson}}\ and\ \bibinfo {author} {\bibfnamefont {S.}~\bibnamefont
  {Redner}},\ }\href@noop {} {\bibfield  {journal} {\bibinfo  {journal}
  {Physical review letters}\ }\textbf {\bibinfo {volume} {114}},\ \bibinfo
  {pages} {198101} (\bibinfo {year} {2015})}\BibitemShut {NoStop}%
\bibitem [{\citenamefont {Majumdar}(2007)}]{majumdar2007brownian}%
  \BibitemOpen
  \bibfield  {author} {\bibinfo {author} {\bibfnamefont {S.~N.}\ \bibnamefont
  {Majumdar}},\ }in\ \href@noop {} {\emph {\bibinfo {booktitle} {The Legacy Of
  Albert Einstein: A Collection of Essays in Celebration of the Year of
  Physics}}}\ (\bibinfo  {publisher} {World Scientific},\ \bibinfo {year}
  {2007})\ pp.\ \bibinfo {pages} {93--129}\BibitemShut {NoStop}%
\bibitem [{\citenamefont {Schuss}\ \emph {et~al.}()\citenamefont {Schuss},
  \citenamefont {Basnayake}, \citenamefont {Guerrier},\ and\ \citenamefont
  {Holcman}}]{Kanishka2}%
  \BibitemOpen
  \bibfield  {author} {\bibinfo {author} {\bibfnamefont {Z.}~\bibnamefont
  {Schuss}}, \bibinfo {author} {\bibfnamefont {K.}~\bibnamefont {Basnayake}},
  \bibinfo {author} {\bibfnamefont {C.}~\bibnamefont {Guerrier}}, \ and\
  \bibinfo {author} {\bibfnamefont {D.}~\bibnamefont {Holcman}},\ }\href@noop
  {} {\bibinfo  {journal} {arXiv preprint arXiv:1711.01330}\ }\BibitemShut
  {NoStop}%
\bibitem [{\citenamefont {Yang}\ \emph {et~al.}(2016)\citenamefont {Yang},
  \citenamefont {Kupka}, \citenamefont {Schuss},\ and\ \citenamefont
  {Holcman}}]{Yang}%
  \BibitemOpen
\bibfield  {journal} {  }\bibfield  {author} {\bibinfo {author} {\bibfnamefont
  {J.}~\bibnamefont {Yang}}, \bibinfo {author} {\bibfnamefont {I.}~\bibnamefont
  {Kupka}}, \bibinfo {author} {\bibfnamefont {Z.}~\bibnamefont {Schuss}}, \
  and\ \bibinfo {author} {\bibfnamefont {D.}~\bibnamefont {Holcman}},\
  }\href@noop {} {\bibfield  {journal} {\bibinfo  {journal} {Journal of
  mathematical biology}\ }\textbf {\bibinfo {volume} {73}},\ \bibinfo {pages}
  {423} (\bibinfo {year} {2016})}\BibitemShut {NoStop}%
\bibitem [{\citenamefont {Holcman}\ and\ \citenamefont
  {Schuss}(2015)}]{holcmanschuss2015}%
  \BibitemOpen
  \bibfield  {author} {\bibinfo {author} {\bibfnamefont {D.}~\bibnamefont
  {Holcman}}\ and\ \bibinfo {author} {\bibfnamefont {Z.}~\bibnamefont
  {Schuss}},\ }\href@noop {} {\bibfield  {journal} {\bibinfo  {journal}
  {Analysis and Applications. Springer, New York}\ } (\bibinfo {year}
  {2015})}\BibitemShut {NoStop}%
\bibitem [{\citenamefont {Aubin}(2013)}]{aubin2013some}%
  \BibitemOpen
  \bibfield  {author} {\bibinfo {author} {\bibfnamefont {T.}~\bibnamefont
  {Aubin}},\ }\href@noop {} {\emph {\bibinfo {title} {Some nonlinear problems
  in Riemannian geometry}}}\ (\bibinfo  {publisher} {Springer Science \&
  Business Media},\ \bibinfo {year} {2013})\BibitemShut {NoStop}%
\bibitem [{\citenamefont {Marchewka}\ and\ \citenamefont
  {Schuss}(2000)}]{marchewka2000path}%
  \BibitemOpen
  \bibfield  {author} {\bibinfo {author} {\bibfnamefont {A.}~\bibnamefont
  {Marchewka}}\ and\ \bibinfo {author} {\bibfnamefont {Z.}~\bibnamefont
  {Schuss}},\ }\href@noop {} {\bibfield  {journal} {\bibinfo  {journal}
  {Physical Review A}\ }\textbf {\bibinfo {volume} {61}},\ \bibinfo {pages}
  {052107} (\bibinfo {year} {2000})}\BibitemShut {NoStop}%
\bibitem [{\citenamefont {Schuss}(2011)}]{schuss2011nonlinear}%
  \BibitemOpen
  \bibfield  {author} {\bibinfo {author} {\bibfnamefont {Z.}~\bibnamefont
  {Schuss}},\ }\href@noop {} {\emph {\bibinfo {title} {Nonlinear filtering and
  optimal phase tracking}}},\ Vol.\ \bibinfo {volume} {180}\ (\bibinfo
  {publisher} {Springer Science \& Business Media},\ \bibinfo {year}
  {2011})\BibitemShut {NoStop}%
\bibitem [{\citenamefont {Freidlin}\ and\ \citenamefont
  {Wentzell}(1998)}]{freidlin1998random}%
  \BibitemOpen
  \bibfield  {author} {\bibinfo {author} {\bibfnamefont {M.~I.}\ \bibnamefont
  {Freidlin}}\ and\ \bibinfo {author} {\bibfnamefont {A.~D.}\ \bibnamefont
  {Wentzell}},\ }in\ \href@noop {} {\emph {\bibinfo {booktitle} {Random
  Perturbations of Dynamical Systems}}}\ (\bibinfo  {publisher} {Springer},\
  \bibinfo {year} {1998})\ pp.\ \bibinfo {pages} {15--43}\BibitemShut {NoStop}%
\bibitem [{\citenamefont {Fain}(1999)}]{Fain}%
  \BibitemOpen
  \bibfield  {author} {\bibinfo {author} {\bibfnamefont {G.~L.}\ \bibnamefont
  {Fain}},\ }\href@noop {} {\emph {\bibinfo {title} {Molecular and cellular
  physiology of neurons}}}\ (\bibinfo  {publisher} {Harvard University Press},\
  \bibinfo {year} {1999})\BibitemShut {NoStop}%
\bibitem [{\citenamefont {Holcman}\ and\ \citenamefont
  {Schuss}(2017)}]{JPA2016-HS}%
  \BibitemOpen
  \bibfield  {author} {\bibinfo {author} {\bibfnamefont {D.}~\bibnamefont
  {Holcman}}\ and\ \bibinfo {author} {\bibfnamefont {Z.}~\bibnamefont
  {Schuss}},\ }\href@noop {} {\bibfield  {journal} {\bibinfo  {journal}
  {Journal of Physics A: Mathematical and Theoretical}\ }\textbf {\bibinfo
  {volume} {50}},\ \bibinfo {pages} {093002} (\bibinfo {year}
  {2017})}\BibitemShut {NoStop}%
\end{thebibliography}%
\end{document}